\documentclass[prl,aps,twocolumn,amsfonts,showpacs,superscriptaddress,preprintnumbers]{revtex4-1}

\usepackage{graphicx}
\usepackage{xcolor}
\usepackage{rotating}
\usepackage{bm,amsmath,amssymb}
\usepackage[utf8]{inputenc}

\usepackage[colorlinks=true, urlcolor=blue, linkcolor=blue, citecolor=blue, hyperindex=true, linktocpage=true]{hyperref}

\def\CA{\mathcal{A}}
\def\CB{\mathcal{B}}

\def\CL{\mathcal{L}}

\def\qfr{\mathfrak{q}}
\def\wfr{\mathfrak{w}}

\def\q{{\bf q}}
\def\x{{\bf x}}

\def\be{\begin{equation}}
\def\ee{\end{equation}}

\def\bea{\begin{eqnarray}}
\def\eea{\end{eqnarray}}

\def\coeff#1#2{{\textstyle {\frac {#1}{#2}}}}

\begin{document}
\preprint{MIT-CTP/5100\;\; OUTP-19-01P}

\title{On the convergence of the gradient expansion in hydrodynamics}

\author{Sa\v{s}o Grozdanov}
\affiliation{Center for Theoretical Physics, MIT, Cambridge, MA 02139, USA}
\author{Pavel K. Kovtun}
\affiliation{Department of Physics \& Astronomy, University of Victoria, PO Box 1700 STN CSC, Victoria,
BC, V8W 2Y2, Canada}
\author{Andrei O. Starinets}
\affiliation{Rudolf Peierls Centre for Theoretical Physics, Clarendon Lab,  Oxford, OX1 3PU, UK}
\author{Petar Tadi\'{c}}
\affiliation{School of Mathematics, Trinity College Dublin, Dublin 2, Ireland}

\begin{abstract}
\noindent
Hydrodynamic excitations corresponding to sound and shear modes in fluids are characterised by gapless dispersion relations. In the hydrodynamic gradient expansion, their frequencies are represented by power series in spatial momenta. We investigate the analytic structure and convergence properties of the hydrodynamic series by studying the associated spectral curve in the space of complexified frequency and complexified spatial momentum. For the strongly coupled ${\cal N}=4$ supersymmetric Yang-Mills plasma, we use the holographic duality methods to demonstrate that the derivative expansions have finite non-zero radii of convergence. Obstruction to the convergence of hydrodynamic series arises from level-crossings in the quasinormal spectrum at complex momenta.
\end{abstract}

\maketitle

{\it Introduction.---}Hydrodynamics is an established universal language for describing near-equilibrium phenomena in fluids~\cite{LL6}. The equations of hydrodynamics are the local conservation laws which can be written as
\be
\label{eq:cons-1}
  \partial_t \rho_a + {\bm\nabla}{\cdot}\,{\bf J}_a = 0,
\ee
where $\rho_a$ are the densities of  locally conserved charges (energy, momentum, particle number, etc.) and ${\bf J}_a$ are the corresponding fluxes. The conservation equations (\ref{eq:cons-1}) can be solved once the fluxes ${\bf J}_a$ are expressed in terms of the densities $\rho_a$ through the so-called constitutive relations, ${\bf J}_a = {\bf J}_a(\rho)$. Conventionally, one works with the quantities $\phi_a$ such as temperature, fluid velocity, and the chemical potential which are conjugate to $\rho_a$ in the grand canonical ensemble. The constitutive relations $\rho_a = \rho_a(\phi)$, ${\bf J}_a = {\bf J}_a(\phi)$ are then used in the conservation laws (\ref{eq:cons-1}) in order to determine the macroscopic space-time evolution of the fluid~\cite{LL6}.

There are two basic physics principles that constrain possible forms of the constitutive relations: symmetry and the derivative expansion. Symmetry is what distinguishes different types of fluids. The derivative expansion is a reflection of the fact that hydrodynamics is only an effective description on length scales much larger than the microscopic scale (such as the mean free path). Thus, the constitutive relations are schematically written as
\begin{subequations}
\label{eq:de-1}
\begin{align}
  & \rho_a = O(\phi) + O(\nabla\phi) + O(\nabla^2\phi) + \dots,\\
  & {\bf J}_a = O(\phi) + O(\nabla\phi) + O(\nabla^2\phi) + \dots,
\end{align}
\end{subequations}
which is a derivative (gradient) expansion. For normal fluids, truncating the expansions (\ref{eq:de-1}) at $O(\phi)$ (i.e.\ neglecting the terms $O(\nabla\phi)$ and higher) gives rise to perfect fluids and Euler equations of hydrodynamics. Truncating the expansions at $O(\nabla\phi)$ gives rise to viscous fluids and Navier-Stokes equations. Truncating at $O(\nabla^2\phi)$ gives rise to second-order hydrodynamics and Burnett equations, and so on.

The naive expectation is that going to higher orders in the derivative expansion improves the hydrodynamic description of the fluid, similar to how the Navier-Stokes equations improve the perfect-fluid approximation by including the viscous effects. The purpose of this paper is thus to address the following foundational question: viewed as an expansion in small gradients, does the hydrodynamic derivative expansion in fact converge?

In order to make this question precise, we will choose a specific physical quantity whose exact value can be compared with the prediction of the derivative expansion. For fluids, the characteristic feature of the hydrodynamic description is the existence of gapless modes: small near-equilibrium fluctuations of the fluid whose frequencies $\omega_i(q)$ are such that $\omega_i(q)\to0$ as the magnitude of the wave vector $q\to0$. The well-known example is the sound wave whose dispersion relation is $\omega_{\rm sound}(q) = \pm v_s q + O(q^2)$, where $v_s$ is the speed of sound. More generally, the hydrodynamic prediction is that 
\be
\label{eq:wq-1}
  \omega_i(q) = \sum_{n=1}^\infty b_n^{(i)} q^n,
\ee
where, in principle, all orders in the derivative expansion (\ref{eq:de-1}) contribute to the dispersion relation of the $i$-th mode. In this paper, we shall investigate whether the infinite power series expansions (\ref{eq:wq-1}) converge, and if so, what determines their radii of convergence. Recently, 
 the convergence of the shear-diffusion mode series in $d=2+1$ was investigated in a holographic model with non-zero chemical potential \cite{Withers:2018srf}. Using the coefficients of the 
series \eqref{eq:wq-1} and Pad\'{e}-approximants, Ref.~\cite{Withers:2018srf} found an obstruction to the convergence in the form of a branch point at purely imaginary momentum and identified this singularity as the collision of two gapped quasinormal modes. Here, we show how branch point singularities generically arise from the spectral curves in classical hydrodynamics.

Besides the general physics interest in the foundations of hydrodynamics, our motivation comes from the success of the relativistic hydrodynamic framework to describe the quark-gluon plasma produced in the collisions of heavy nuclei~\cite{Gale:2013da}. Similarly, in the examples of strongly interacting quantum field theories whose non-equilibrium evolution can be determined from first principles using holographic methods, the hydrodynamic description appears to be unexpectedly robust, even when the gradients are large~\cite{Chesler:2009cy,Heller:2011ju}. If the expansions (\ref{eq:wq-1}) indeed converge, this convergence would be a step towards understanding the ``unreasonable effectiveness'' of the hydrodynamic description of the quark-gluon plasma and of similar strongly interacting holographic fluids \cite{Romatschke:2017ejr,Busza:2018rrf,Buchel:2016cbj,Grozdanov:2016vgg}.

An important comment has to be made before we proceed. The expansion (\ref{eq:wq-1}) is a prediction of classical hydrodynamics, which neglects the effects of statistical fluctuations. As is well known~\cite{Ernst1975},  fluctuations lead to infinitely many fractional powers of $q$ appearing in $\omega_i(q)$, thereby rendering the expansion~(\ref{eq:wq-1}) insufficient. While one may rightly question the applicability of classical second- and higher-order hydrodynamics to the quark-gluon plasma on these grounds~\cite{Kovtun:2011np}, one should keep in mind that the complete effective description of the fluid involves both the classical hydrodynamics and the fluctuation effects. Our focus here is on the classical hydrodynamics part, with the understanding that the fluctuation effects are to be included later. In holographic hydrodynamics, these effects  are suppressed in the large-$N$ limit of the corresponding models~\cite{Kovtun:2003vj}.

{\it Hydrodynamic modes from complex curves.---}In order to understand the origin of the series (\ref{eq:wq-1}), consider the constitutive relations (\ref{eq:de-1}) truncated at a finite order $k$ in the derivative expansion. Suppose there are $m$ hydrodynamic variables $\phi_a$, with $a=1,\ldots,m$. The hydrodynamic equations (\ref{eq:cons-1}) linearised near equilibrium can be written as $\CL_a[\delta\phi] = 0$, where $\CL_a$ are linear differential operators of order at most $k{+}1$. Upon Fourier transforming the linearised fluctuations, $\delta\phi \propto \exp(-i\omega t+ i\q{\cdot}\x)$, the frequencies are determined by the eigenvalue equation $\det L_{ab}(\omega,\q)=0$, where $L_{ab}$ is an $m\times m$ matrix whose elements are polynomials in $\omega$ and $\q$. Assuming a rotation-invariant equilibrium state, the eigenvalue equation can only depend on $\q^2$ and $\omega$, and can be written as $P_k(\q^2, \omega) = 0$, where $P_k$ is a polynomial in  $\q^2$ and $\omega$.  Continuing the gradient expansion, we set $P(\q^2, \omega ) = \lim_{k\rightarrow \infty} P_k(\q^2, \omega)$. The all-order eigenvalue equation is then
\be
\label{eq:Pwq-1}
  P(\q^2, \omega) = 0.
\ee
It is useful to treat $z\equiv \q^2$ and $\omega$ as complex variables. The eigenvalue equation $P(z,\omega) = 0$ then defines a complex spectral curve in $\mathbb{C}^2$. Regular points of the curve satisfy the condition 
of the analytic  implicit function theorem \cite{gunning-rossi},
\begin{align}
P(z, \omega )=0, \qquad \frac{\partial P(z, \omega )}{\partial \omega} \neq 0,
\label{c-curve-regular}
\end{align}
which guarantees analyticity and uniqueness of the  branch $\omega=\omega(z)$ in the vicinity of a regular point. Of particular interest are the so-called critical points, i.e. the points $(z_{\rm c},\omega_{\rm c})$ where, in addition to $P(z_{\rm c}, \omega_{\rm c} )=0$, the first $(p{-}1)$ derivatives with respect to $\omega$ vanish,
\begin{align}
 \frac{\partial P(z_{\rm c}, \omega_{\rm c} )}{\partial \omega} =0, \, \ldots, \,
\frac{\partial^{p} P(z_{\rm c}, \omega_{\rm c} )}{\partial \omega^{p}} \neq 0 .
\label{c-curve-critical}
\end{align}
Assuming the analyticity of $P(z, \omega)$ at $(z_{\rm c},\omega_{\rm c})$, the generalisation of the implicit function theorem guarantees the existence of $p$ branches $\omega_j=\omega_j(z)$, $j=1,\ldots,p$, represented by Puiseux series (series in fractional powers of $(z-z_{\rm c})$) converging in the vicinity of the branch point $z=z_{\rm c}$ \cite{Wall-singular-points}. 
The hydrodynamic (gapless) dispersion relations are defined implicitly by Eq.~(\ref{eq:Pwq-1}). They arise as $m$ functions $\omega_i=\omega_i(z)$ satisfying $\omega_i(z\to0)=0$, $i=1,\ldots,m$, and $P(z, \omega(z))=0$.

{\it Relativistic hydrodynamics.---}In what follows, we focus on relativistic hydrodynamics for concreteness~\cite{Kovtun:2012rj}, and set  $\hbar=c=1$. In first-order hydrodynamics of an uncharged fluid in 3+1 dimensions, one finds
\be
\label{eq:Pwq-1x}
  P_1(\q^2, \omega) = \left( \omega + i D  \q^2 \right)^2 \left(\omega^2 + i  \Gamma \omega  \q^2 - v_s^2   \q^2  \right) = 0,
\ee
where $v_s = (\partial p/\partial\epsilon)^{1/2}$ is the speed of sound, $p$ and $\epsilon$ are the equilibrium pressure and energy density, $D = \eta/(\epsilon{+}p)$ is the diffusion coefficient of the transverse velocity, $\Gamma = (\coeff 43 \eta + \zeta)/(\epsilon+p)$ is the damping coefficient of sound waves,  $\eta$ is the shear viscosity and $\zeta$ is the bulk viscosity. 
More generally, one can show that the eigenvalue equation (\ref{eq:Pwq-1}) factorises as $P(\q^2,\omega)=F_{\rm shear}^2 F_{\rm sound}$, with
\begin{align}
  F_{\rm shear} & \equiv \omega + i\q^2 \gamma_\eta(\q^2,\omega) = 0 , 
\label{shear-cc}  \\
  F_{\rm sound} & \equiv \omega^2 + i\omega\q^2 \gamma_s(\q^2,\omega) - \q^2 H(\q^2,\omega) =0.
\label{sound-cc}
\end{align}
This factorisation is a consequence of rotation invariance. In the derivative expansion, the functions $\gamma_\eta$, $\gamma_s$ and $H$ are given by power series around  $(z, \omega)= (0,0)$. 

For example, assuming analyticity of the function $\gamma_\eta$ at $(z, \omega) =(0,0)$, the origin $(0,0)$  is a regular point of the complex curve \eqref{shear-cc}. The implicit function theorem then implies that the shear dispersion relation is given by a series in powers of $\q^2$ converging in the vicinity of $\q^2=0$,
\begin{align}
\label{eq:shear-1}
   \wfr_{\rm shear} =  -i \sum_{n=1}^{\infty} c_n \qfr^{2 n} = -i\, c_1 \qfr^2 + \ldots ,
\end{align}
where we defined $\wfr \equiv \omega/2\pi T$, $\qfr \equiv |\q|/2\pi T$ and $c_1 = 2\pi T D$. On the other hand, for the curve \eqref{sound-cc} we have 
$\partial F_{\rm sound} /\partial \omega =0$ but $\partial^2 F_{\rm sound} /\partial \omega^2 \neq 0$ at $(0,0)$---again, assuming analyticity of the functions $\gamma_s$ and $H$ at the origin. In this case, the two branches $\omega^\pm =\omega^\pm (\q^2)$ are given by Puiseux series in powers of $q\equiv (\q^{2})^{\frac{1}{2}}$,
\begin{align}
\label{eq:sound-1}
  \wfr_{\rm sound}^\pm = -i  \sum_{n=1}^{\infty} a_n e^{\pm \frac{i \pi n}{2}} \qfr^{n} =  
\pm a_1 \qfr + i a_2 \qfr^2 + \ldots,
\end{align}
converging in the vicinity of $\q^2=0$. The dimensionless coefficients are $a_1=v_s$, $a_2 = -\Gamma\pi T$. We expect the radius of convergence of the hydrodynamic dispersion relation series to be 
determined by the distance from the origin to the nearest critical point \eqref{c-curve-critical}.

\begin{figure}
 \includegraphics[width=0.45\textwidth]{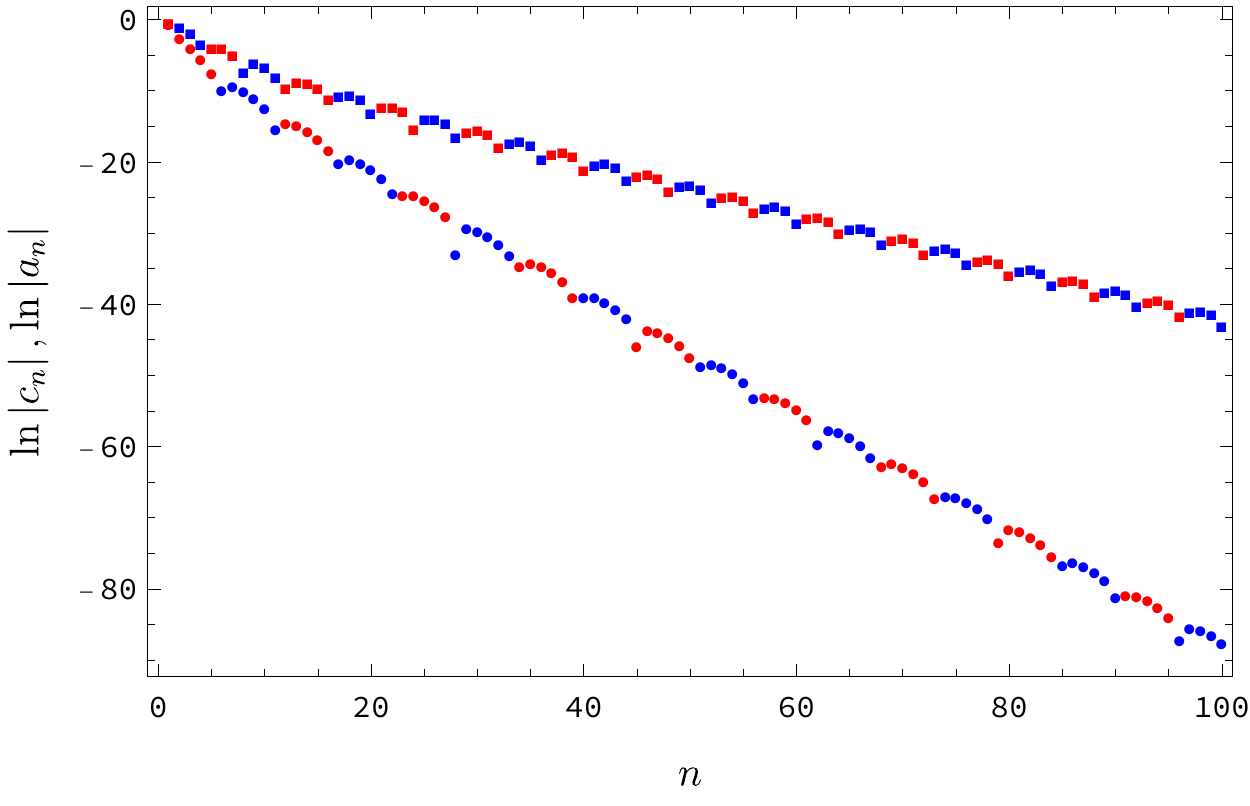}
\caption{
\label{fig:log-cn-shear-sound}
Coefficients of the expansions~\eqref{eq:shear-1} and \eqref{eq:sound-1} in ${\cal N}=4$ SYM theory. The circles are $\ln|c_n|$ (shear mode), the squares are $\ln|a_n|$ (sound mode). Red (blue) colour indicates positive (negative) 
values of $c_n$ or $a_n$.
}
\end{figure}
{\it The main example.---}An example of a  quantum field theory in which the dispersion relations $\omega_i(q)$ can be analysed to all orders in $q$ is the ${\cal N} = 4$ supersymmetric $SU(N)$ Yang-Mills (SYM) theory at infinitely large  't Hooft coupling and infinite 
$N$. This theory has been extensively studied through the use of holographic methods in various contexts \cite{Aharony:1999ti,Ammon:2015wua}, including as a model of collective properties of quantum chromodynamics in the deconfined phase~\cite{CasalderreySolana:2011us}. The theory is conformal, and the only dimensionful scale is the equilibrium temperature $T$ that sets the finite correlation length. 
In any conformal theory, we have $a_1 = 1/\sqrt{3}$, $a_2=-\coeff23 c_1$, and the bulk viscosity $\zeta=0$. In 
 ${\cal N} = 4$ SYM theory, we further have $c_1 = 1/2$ due to the universal holographic relation $\eta/s=1/4\pi$ (where $s$ is the density of entropy)~\cite{Kovtun:2004de}.

{\it Holography.---}To compute the coefficients $c_n$ and $a_n$ in ${\cal N} = 4$ SYM theory, we 
use the holographic duality to map the quantum field-theoretic problem into a calculation in classical general relativity. The duality implies that the hydrodynamic modes \eqref{eq:wq-1} coincide with the gapless quasinormal modes of black branes in one higher dimension \cite{Son:2002sd,Kovtun:2005ev}. The relevant gravitational perturbations of the black brane are described by two functions: $Z_1(u)$ (the shear mode) and $Z_2(u)$ (the sound mode), where $u$ is the radial coordinate ranging from $u=0$ (asymptotic boundary) to $u=1$ (event horizon) \cite{Kovtun:2005ev}. The shear mode equation is
\be
    Z_1'' - \frac{(\wfr^2 - \qfr^2 f)f - u\wfr^2 f'}{uf(\wfr^2-\qfr^2 f)} Z_1'
    + \frac{\wfr^2 - \qfr^2 f}{u f^2} Z_1 = 0,
\label{eq:Z1eqn}
\ee
where $f(u)=1-u^2$. The sound mode equation is 
\begin{widetext}
\be
    Z_2'' - 
    \frac{3\wfr^2 (1+u^2) + \qfr^2 ( 2u^2 - 3 u^4 -3)}
         {u f (3 \wfr^2 +\qfr^2 (u^2-3))} \, Z_2'
    + \frac{3 \wfr^4 +\qfr^4 ( 3-4 u^2 + u^4) +
    \qfr^2 ( 4 u^5 - 4 u^3 + 4 u^2 \wfr^2 - 6 \wfr^2)}
    {u f^2 ( 3 \wfr^2 + \qfr^2 (u^2 -3))}\, Z_2
    = 0.
\label{eq:Z2eqn}
\ee
\end{widetext}
Both equations have to be solved with the boundary condition $Z_i(u)\sim(1-u)^{-i\wfr/2}$ as $u\to1$, corresponding to the infalling wave at the horizon. Near the boundary $u=0$, the two independent solutions have exponents $0$ and $2$. Hence, the solution satisfying the infalling condition at the horizon can be written near the boundary as $Z_i(u)\sim \CA_i (1+\dots) + \CB_i u^2+ \dots$, where the dots denote higher powers of $u$, and $\CA_i$, $\CB_i$ are the two integration constants which depend on $\wfr$ and $\qfr^2$. The Dirichlet condition (the holographic analogue of Eqs.~\eqref{shear-cc} and \eqref{sound-cc}),
\be
\label{eq:A0}
  \CA_i(\qfr^2,\wfr) = 0,
\ee
relates $\wfr$ to $\qfr^2$ and gives the dispersion relations of the hydrodynamic and other (gapped) modes~\cite{Kovtun:2005ev}. We note that the analyticity of the coefficient $\CA_i(\qfr^2,\wfr)$ at the origin is not {\it a priori} guaranteed, and thus it is not obvious that the hydrodynamic series in holography have a non-zero radius of convergence. In order to find the coefficients $c_n$, $a_n$ in the expansions~\eqref{eq:shear-1}, \eqref{eq:sound-1}, we must solve Eqs.~(\ref{eq:Z1eqn}), (\ref{eq:Z2eqn}). We do this by constructing the Frobenius series solution at $u=1$, and truncating the series at a sufficiently high order \cite{Kovtun:2005ev}.

\begin{figure*}[t]
\centering
\includegraphics[width=0.325\textwidth]{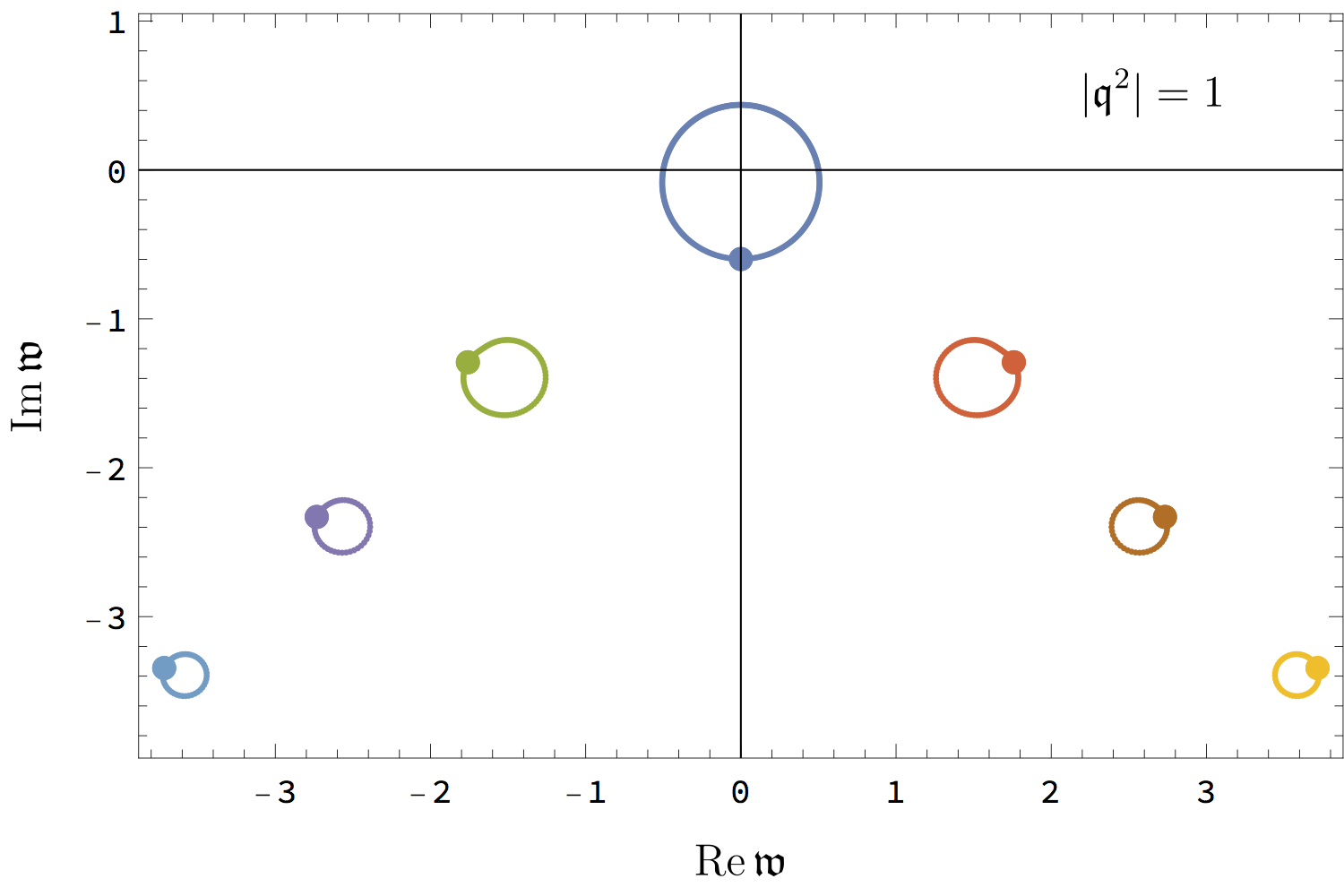}
\includegraphics[width=0.325\textwidth]{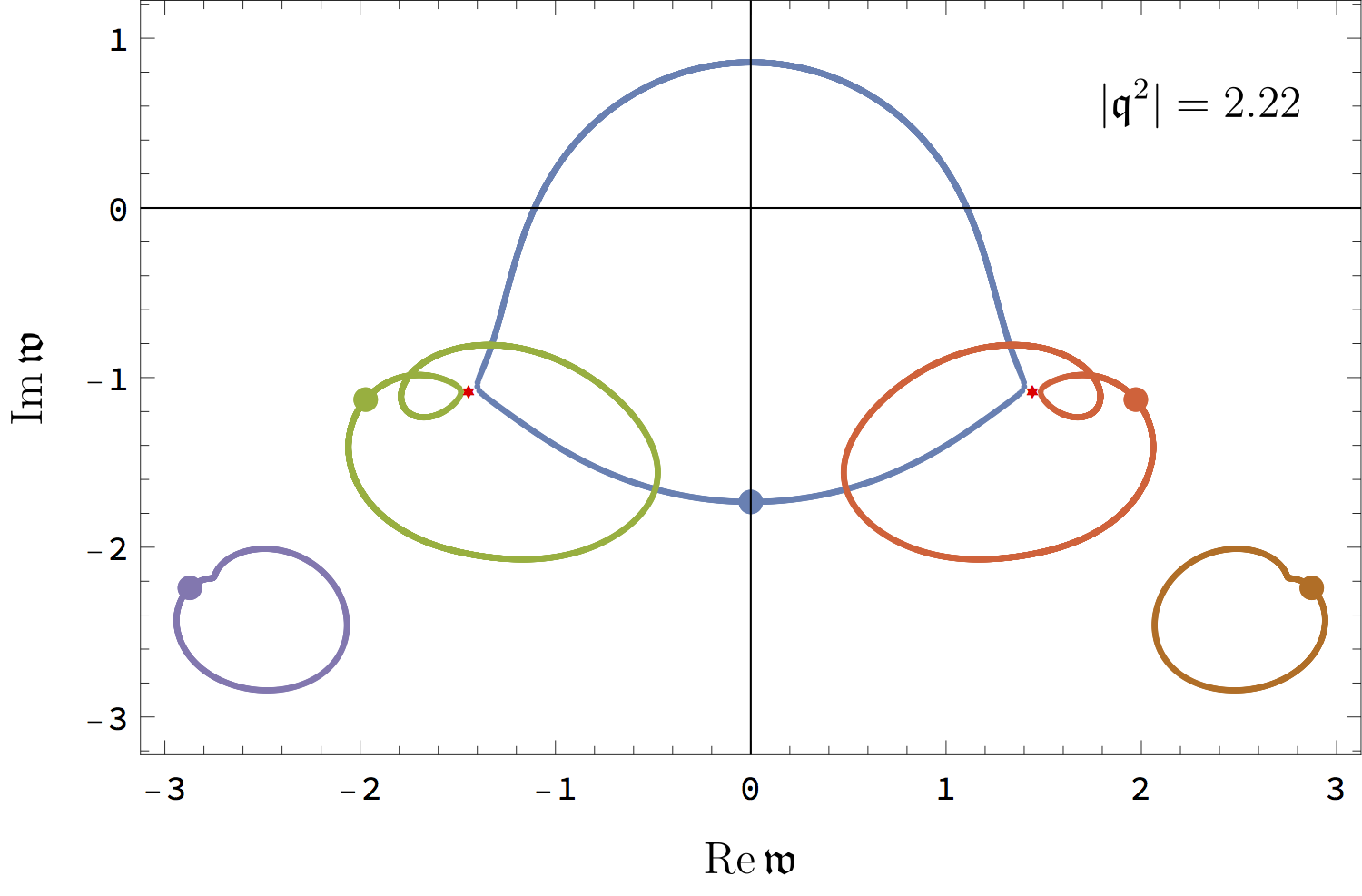}
\includegraphics[width=0.325\textwidth]{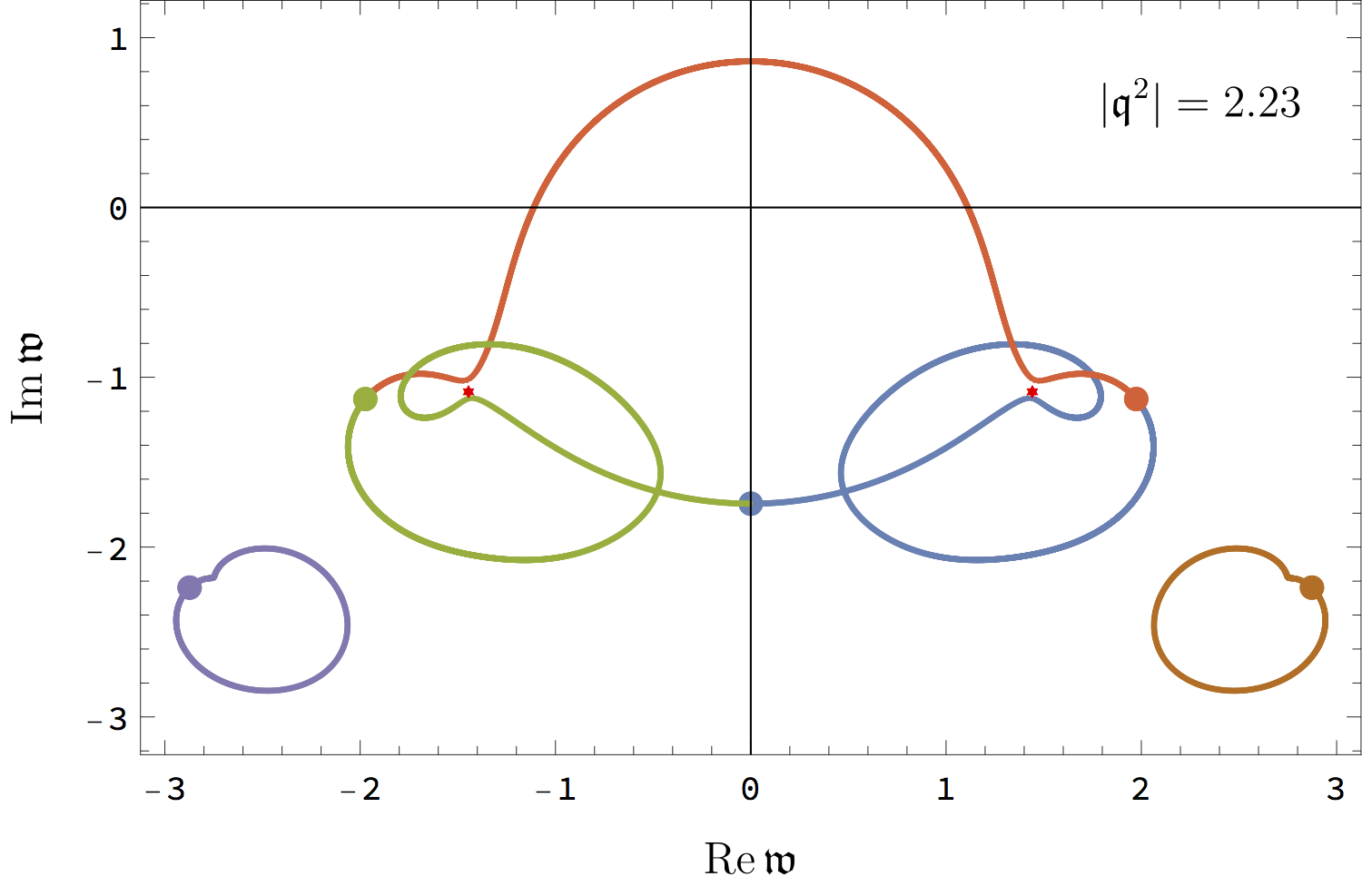}
\\
\includegraphics[width=0.325\textwidth]{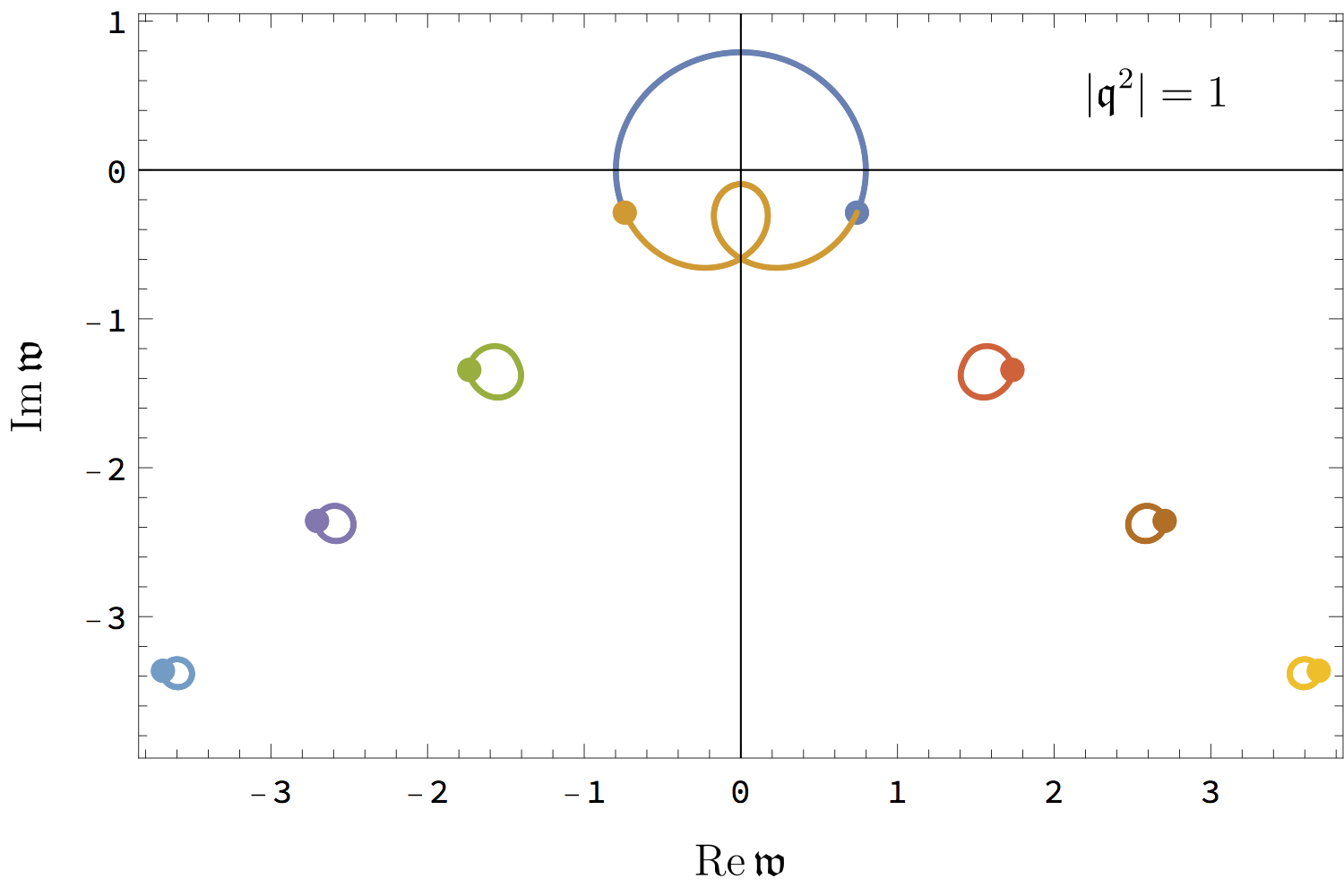}
\includegraphics[width=0.325\textwidth]{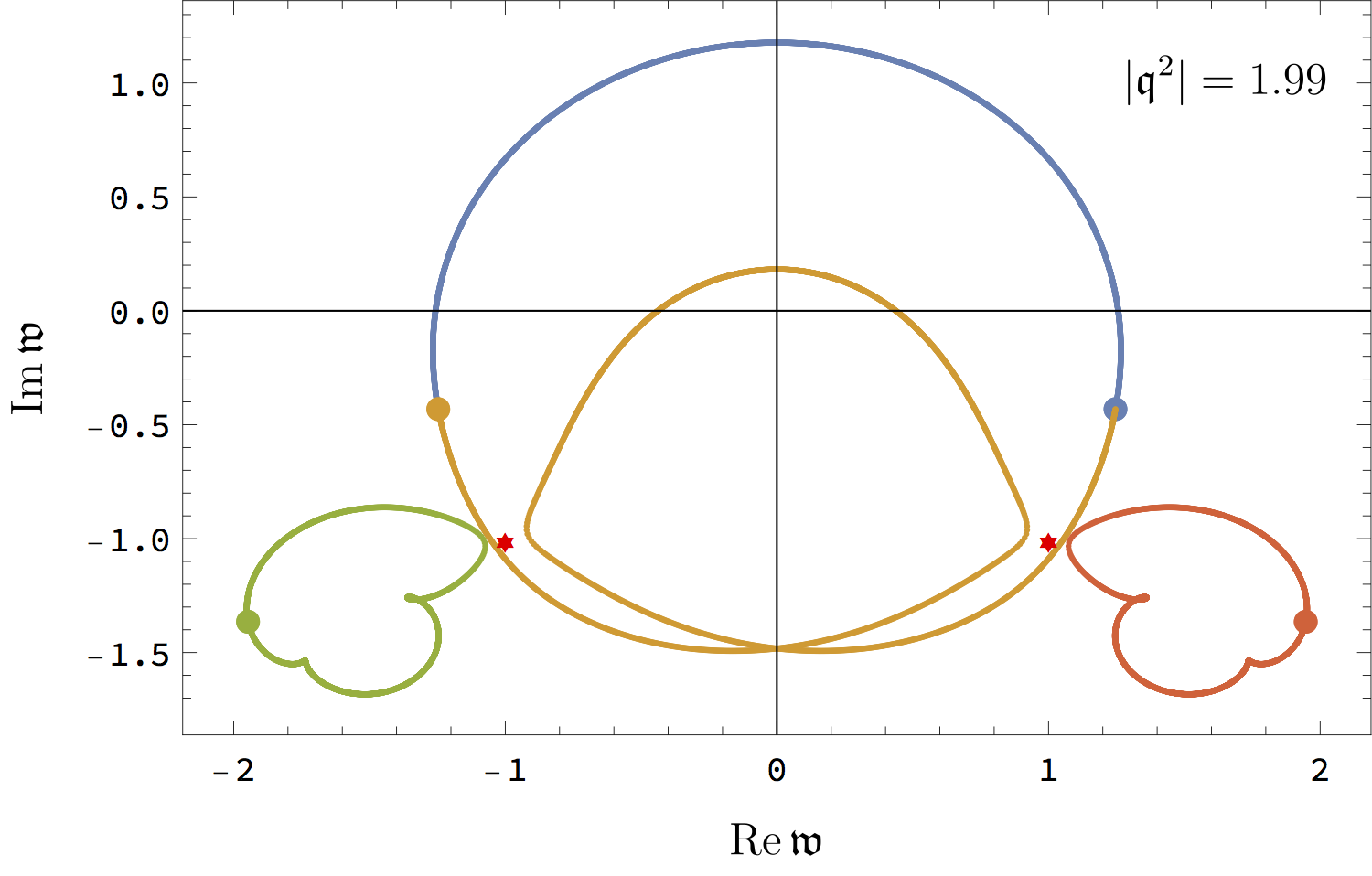}
\includegraphics[width=0.325\textwidth]{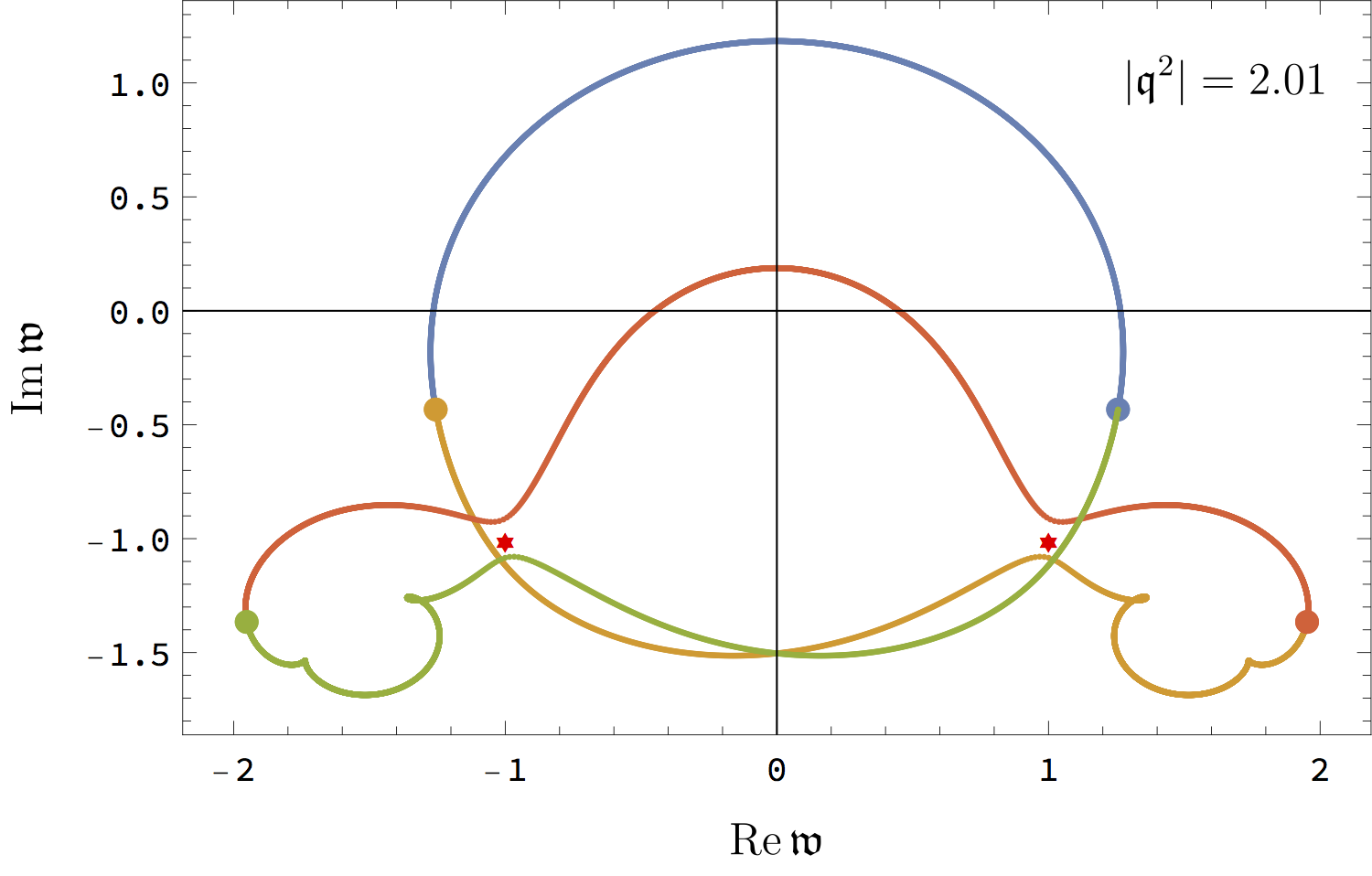}
\caption{
\label{fig:pole-collisions-complex-q}
Poles of the retarded two-point function of the energy-momentum tensor in the complex $\wfr$-plane, at various values of the complexified momentum $\qfr^2 = |\qfr^2|e^{i \theta}$. Top row is the shear channel, bottom row is the sound channel. Large dots correspond to the location of the poles for real $\qfr^2$ ($\theta=0$) \cite{Kovtun:2005ev}. The hydrodynamic shear and sound poles are the poles closest to the real axis in the top left and bottom left panels, correspondingly. As $\theta$ increases from $0$ to $2\pi$, each pole moves counter-clockwise, following the trajectory of its colour. In the shear channel (top row), at $|\qfr^2|=1$, each pole follows a closed orbit (top left). At $|\qfr^2|=2.22$ (top centre), the hydrodynamic pole almost collides with the two gapped poles closest to the real axis. The actual collision would happen at the critical momentum (\ref{eq:qw-crit-shear}), $|\qfr_{\rm c}^2| \approx 2.224$, with the corresponding frequencies marked by red asterisks in the figure.  At $|\qfr^2|=2.23$ (top right), the orbits of the three uppermost poles are no longer closed: the hydrodynamic pole and the two gapped poles exchange their positions cyclically as the phase $\theta$ increases from $0$ to $2\pi$. Similar behaviour is observed for the sound mode (bottom row). The dispersion relations $\wfr_i(\qfr)$ thus have branch cuts starting at  $\qfr_{\rm c}$.
}
\end{figure*}

The results are shown in Fig.~\ref{fig:log-cn-shear-sound}. The plots indicate that the coefficients $c_n$ and $a_n$ decrease exponentially with $n$, and therefore the convergence radii  are non-zero for both the shear and the sound modes. Finite radii of convergence of the series~\eqref{eq:shear-1} and \eqref{eq:sound-1} imply the existence of singularities in the complex $\qfr$-plane obstructing analyticity. The absolute value of the critical $\qfr^{\rm c}_i$ that sets the finite radius of convergence is determined by the slope, and the argument of $\qfr^{\rm c}_i$ is determined by the period of the oscillations in Fig.~\ref{fig:log-cn-shear-sound}.
While the critical values $\qfr^{\rm c}_i$ can be extracted from these data by fitting the coefficients to exponential functions of $n$ with complex exponents, a more precise way to find $\qfr_i^{\rm c}$ is by solving the set of equations (the holographic analogue of Eq.~\eqref{c-curve-critical})
\begin{align}
\label{eq:AdA-1}
  \CA_i(\qfr^2,\wfr) = 0, \qquad \frac{\partial \CA_i(\qfr^2, \wfr)}{\partial\wfr} = 0,
\end{align}
which determines the critical points. The Frobenius expansion around the horizon gives $\CA_i(\qfr^2, \wfr)$ as explicit algebraic functions of $\wfr$ and $\qfr^2$, and Eqs.~(\ref{eq:AdA-1}) can be solved numerically. For the shear mode, we find two pairs $(\qfr_{\rm c}^2, \wfr_{\rm c})$ with
\begin{subequations}
\label{eq:qw-crit-shear}
\begin{align}
   \qfr_{\rm c}^2 & \approx 1.8906469 \pm 1.1711505 i,\\
   \wfr_{\rm c} & \approx \pm 1.4436414 - 1.0692250 i,
\end{align}
\end{subequations}
corresponding to the convergence radius of the shear mode dispersion relation $|\qfr_{\rm shear}^{\rm c}| \approx 1.49131$. For the sound mode, we find, similarly,
\begin{align}
\label{eq:qw-crit-sound}
   \qfr_{\rm c}^2 = \pm 2i, \qquad \wfr_{\rm c} = \pm 1 - i ,
\end{align}
within the limits of our numerical accuracy. One can check that the values (\ref{eq:qw-crit-sound}) indeed satisfy Eq.~(\ref{eq:A0}), with a simple analytic solution for $Z_2(u)$. This corresponds to the convergence radius of the sound mode dispersion relation 
$|\qfr_{\rm sound}^{\rm c}| = \sqrt{2} \approx 1.41421$. 
The values of $\qfr_{\rm c}$ for the shear and sound modes are not equal, but are quite close.
Thus the slopes of the two lines in Fig.~\ref{fig:log-cn-shear-sound} differ by approximately a factor of 2, as the shear mode frequency is expanded in $\qfr^2$, while the sound mode frequency is expanded in $(\qfr^2)^{1/2}$.

{\it Quasinormal spectrum level-crossing.---}The origin of the critical values (\ref{eq:qw-crit-shear}) and (\ref{eq:qw-crit-sound}) can be understood if we consider the singularities (poles) of the retarded two-point correlation functions of the energy-momentum tensor for complex $\wfr, \qfr$. The poles, or quasinormal frequencies, determined by Eq.~(\ref{eq:A0}), contain the hydrodynamic modes $\wfr_i(\qfr)$ as well as an infinite tower of gapped modes $\wfr^{\rm gapped}_n(\qfr)$, such that $\wfr^{\rm gapped}_n(\qfr \to 0)\neq0$ \cite{Kovtun:2005ev}. Consider now the locations of the poles $\wfr(\qfr)$ in the complex $\wfr$ plane as the phase $\theta$ of the complex momentum $\qfr^2 = |\qfr^2|e^{i\theta}$ changes from $0$ to $2\pi$, as illustrated in Fig.~\ref{fig:pole-collisions-complex-q}. At small $|\qfr^2|$, the poles (whose original location at $\theta=0$ is indicated by large dots) move along simple curves, as shown in the left-most panels of Fig.~\ref{fig:pole-collisions-complex-q}. With $|\qfr^2|$ increasing, the trajectories of the poles exhibit more complicated behaviour. The poles effectively ``interact'' with each other, and at special values of $\qfr=\qfr_{\rm c}$, the hydrodynamic poles collide with one of the gapped-mode poles. This is illustrated in Fig.~\ref{fig:pole-collisions-complex-q}, where we show the trajectories of the poles just before and just after the collision, with the collision points marked by asterisks. The figures clearly show that the collision of poles happens at the critical values given by  Eqs.~(\ref{eq:qw-crit-shear}), (\ref{eq:qw-crit-sound}) when the  hydrodynamic poles transform into one of the  former gapped poles. With $|\qfr^2|$ further increasing, other poles from the infinite tower of gapped 
quasinormal modes become involved. By analogy with quantum mechanics, we call this phenomenon the quasinormal  spectrum level-crossing.  Thus, the radius of convergence  of the hydrodynamic series $|\qfr_{\rm c}|$ can be viewed as the absolute value of (complex) $\qfr$ with the smallest possible $| \qfr |$ at which the hydrodynamic pole collides with a gapped pole.

{\it Discussion.---}We have shown that the gradient expansions for the hydrodynamic shear and sound frequencies in the strongly coupled ${\cal N} = 4$ SYM theory have finite radii of convergence given by $q_{\rm sound}^{\rm c} = \sqrt{2}\, (2 \pi T)$ for the sound mode, and by $q_{\rm shear}^{\rm c}\approx 1.49\, (2 \pi T)$ for the shear mode. In general, all-order hydrodynamics gives rise to convergent dispersion relations \eqref{eq:wq-1}, provided the analyticity of the corresponding spectral curves at the origin is established independently. While the radius of convergence could, in principle, be infinite, in the example of ${\cal N} = 4$ SYM theory, it was limited by the collision of the poles of the two-point correlation function of the energy-momentum tensor at complex $q$. This obstruction to convergence would be invisible had we only considered real values of $q$.

Returning to the question of the ``unreasonable effectiveness'' of hydrodynamics, we note that the derivative expansion in relativistic hydrodynamics has  been previously argued to diverge~\cite{Heller:2013fn}. This is based on assuming that the fluid undergoes a one-dimensional expansion, such that all quantities only depend 
on proper time $\tau$. One assumes that an expansion of the fluid energy density in powers of $\tau^{-2/3}$ can be performed, identifying this as a gradient expansion. This large-$\tau$ expansion is divergent in holographic models~\cite{Heller:2013fn}, and in the M\"{u}ller-Israel-Stewart (MIS) extension of hydrodynamics~\cite{Heller:2015dha}. The gradient expansion we are considering here is different and thus our results do not contradict \cite{Heller:2013fn,Heller:2015dha}. Our interest is in the near-equilibrium spatial gradient expansion, rather than in the boost-invariant flow. In the same MIS theory where the large-$\tau$ expansion diverges, the small-$q$ expansions of Eq.~(\ref{eq:wq-1}) converge \cite{Baier:2007ix}. Here we have shown that in the same strongly coupled ${\cal N}=4$ SYM theory where the large-$\tau$ expansion diverges, the small-$q$ expansions of Eq.~(\ref{eq:wq-1}), again, converge. Therefore one must be careful not to identify the divergence of the large-$\tau$ expansion with the failure of the hydrodynamic gradient expansion. The common conclusion of our results and those of \cite{Heller:2013fn} is that the presence of the non-hydrodynamic degrees of freedom (the gapped quasinormal modes) is what sets the limit on the applicability of  hydrodynamics. 

Finally, we comment on the dependence of the radii of convergence on coupling, limiting the discussion to the sound mode in first-order
hydrodynamics, where the first non-trivial critical 
point occurs at $|\qfr_{\rm sound}^{\rm c}| = v_s/\pi T \Gamma$. For conformal theories $v_s=1/\sqrt{d}$, and at infinitely strong coupling, dual gravity approximation gives $|\qfr_{\rm sound}^{\rm c}| =2 \sqrt{d}/(d-1) =\sqrt{3}$ in $d=3$, not too far from the correct value  in Eq.~\eqref{eq:qw-crit-sound}. For ${\cal N}=4$ SYM theory, taking into 
account the leading order correction to the shear viscosity - entropy density ratio at large but finite 't Hooft coupling $\lambda$ \cite{Buchel:2004di}, we find 
$|\qfr_{\rm sound}^{\rm c}| = \sqrt{3} \left( 1 - 15\zeta (3) \lambda^{-3/2} + \cdots\right)$. Therefore, it appears that the radius of convergence is smaller at weaker coupling, in line with the earlier observations regarding validity of hydrodynamics at finite coupling \cite{Grozdanov:2016vgg,Grozdanov:2016zjj}.

{\it Acknowledgements.---}S.G. was supported by the U.S. DOE grant  DE-SC0011090.   P.K. was supported in part by NSERC of Canada.  The work of P.T. is supported by an Ussher Fellowship from Trinity College Dublin.

\bibliographystyle{apsrev4-1}
\bibliography{hydrostuff}{}

\end{document}